\documentstyle[12pt]{article}
\begin{document}

\thispagestyle{empty}
\rightline{NSF-ITP-94-83}
\rightline{UCSBTH-94-30}
\rightline{hep-th/9408134}

\vskip 2.5 cm

\begin{center}
{\large\bf  Hot String Soup }\break

\vskip 1.0 cm

{\bf David A. Lowe} \footnote{Electronic address:
lowe@tpau.physics.ucsb.edu}

\vskip 5pt

{\sl Department of Physics \break
University of California \break
Santa Barbara, CA 93106-9530} \break

and {\bf L{\'a}rus Thorlacius} \footnote{Electronic address:
larus@nsfitp.itp.ucsb.edu}

\vskip 5pt

{\sl Institute for Theoretical Physics  \break
University of California  \break
Santa Barbara, CA 93106-4030} \break

\end{center}

\vskip 1.0 cm

\begin{quote}

Above the Hagedorn energy density closed fundamental strings
form a long string phase.  The dynamics of weakly interacting
long strings is described by a simple Boltzmann equation which
can be solved explicitly for equilibrium distributions.  The
average total number of long strings grows logarithmically with
total energy in the microcanonical ensemble.  This is consistent
with calculations of the free single string density of states
provided the thermodynamic limit is carefully defined.  If the
theory contains open strings the long string phase is suppressed.

\end{quote}

\newpage

\section{Introduction}

The statistical mechanics of strings at high temperature
differs significantly from that of pointlike objects.  The
exponential growth in the single-string density of states as a
function of mass \cite{hage}-\cite{carl} makes the
canonical partition function,
\begin{equation}
\label{canpart}
Z={\rm tr}\, [e^{-\beta H}],
\end{equation}
of a free string gas ill-defined as one approaches the so-called
Hagedorn temperature.  Different physical interpretations of this fact
have been offered, including that the Hagedorn temperature defines an
absolute limiting temperature in physics \cite{hage}-\cite{huang}
or that it signals a transition to an unknown high-temperature
phase where strings may be replaced by more fundamental degrees of
freedom \cite{atick}.

The high energy limit of the single string density of states for free
closed strings in $D$ non-compact space dimensions is found
to be \cite{gsw}
\begin{equation}
\omega(\varepsilon) \approx
{V\,\exp{(\beta_H\varepsilon)}\over \varepsilon^{D/2+1}} ,
\label{freedos}
\end{equation}
where $V$ is the $D$-volume occupied by the system and $\beta_H$ is the
inverse Hagedorn temperature.
Given this density of states one can compute the energy
distribution function $d(\varepsilon,E)$ in the microcanonical ensemble,
which gives the average number of
strings carrying energy between $\varepsilon$
and $\varepsilon+\delta\varepsilon$ in a system of total energy $E$,
and at high energy density it is found to favor the formation of a
single long string which carries most of the available
energy \cite{fraut},\cite{carl},\cite{sund}-\cite{deo}.

This physical picture is suspect because it neglects the effect of
interactions.  A long string, which carries a finite fraction of the
total energy in a system at high energy density, will extend further
than the size of the system itself and as the thermodynamic limit is
approached it
traverses the entire volume of the system many times over.  In this
limit the string has numerous opportunities to intersect itself and
it is natural to ask how string interactions affect the equilibrium
configuration.  We address this question by writing down a Boltzmann
equation appropriate for very long closed strings and obtaining
self-consistent solutions.  In the high energy limit the density of
states takes the form
\begin{equation}
\omega(\varepsilon) \approx
{\exp{(\beta_H \varepsilon)} \over \varepsilon},
\label{intdos}
\end{equation}
instead of (\ref{freedos}) and this has important consequences for
the energy distribution function and the thermodynamic behavior of
the system.  There is still a long string phase
but it is dominated by a large number of long strings which may
split and join.  Similar conclusions were reached by
Salomonson and Skagerstam \cite{salom} in the context of a discrete
model of strings, but these seem to have been
overlooked in much of the subsequent literature on the subject.

The discrepancy between (\ref{freedos}) and (\ref{intdos}) can be
understood as follows.
The microcanonical ensemble is defined only for finite
volume systems with finite total energy and some care is required
in defining the thermodynamic limit for extended objects such as
strings.  At present the only consistent description of strings in
finite volume is to consider a compact target space.  The density
of states in the thermodynamic limit is then obtained by performing
a calculation on a finite sized target space and then letting the
size tend to infinity at the end of the day.

There is a further subtlety which needs to be considered.
The assumption of equipartition, which is needed to make contact
between the density of states and equilibrium distribution functions,
requires the existence of interactions, however weak they may be.
Since string interactions always include gravity the thermal ensemble
is only defined for length scales satisfying
\begin{equation}
\label{jeancond}
R^2 < {1\over g^2 \rho} ,
\end{equation}
due to the Jeans instability.  In the thermodynamic limit we wish to
consider fixed energy density, $\rho$, as the volume becomes large and
from (\ref{jeancond}) we see that the limit can only be taken if at the
same time the string coupling, $g$, is scaled to zero sufficiently
rapidly.  Alternatively, we can at the outset fix the value of the
string coupling to be extremely small and then restrict our
considerations to systems of large, but finite, volume consistent
with (\ref{jeancond}).

Salomonson and Skagerstam \cite{salom} showed that when the extended
dimensions of space are taken to be compactified on a $D$-torus the
free string density of states for closed strings indeed takes the form
(\ref{intdos}) rather than (\ref{freedos}), which was obtained for
strings in non-compact space, and this was later confirmed by other
groups \cite{bowgid},\cite{deo}.  One might worry that
this result depends on the choice of topology that is used to implement
string theory in finite volume.  For example, one could consider
instead of a torus a group manifold of some simple group where
strings have no winding modes and the counting of high energy states
is naively quite different from the toroidal case.  It turns out,
however, that the global topology plays no role.  A general
argument based on modular invariance, given by Brandenberger and
Vafa \cite{branden}, shows that the free density of states for closed
strings takes the form (\ref{intdos}) for sufficiently high energy
on any compact target space.

Thus the very existence of interactions has a subtle but important
effect on the density of states of strings at high energy.  This effect
is entirely due to the extended nature of strings and can be seen in
two different ways.  On the one hand the Jeans instability forces one
to consider only strings on a compact target space and this affects
the counting of free single string states.  The other method will be
discussed in the present paper and involves solving a set of transport
equations for weakly interacting strings.  These equations are quite
simple in the long string limit and relatively little work is required
to get at the density of single string states using our approach.

The plan of this paper is as follows. In Section~\ref{secii} we
introduce the Boltzmann equation for long closed strings and obtain
self-consistent equilibrium solutions. These solutions agree with the
discrete string model results of Salomonson and Skagerstam \cite{salom}.
The Boltzmann equation also allows one to consider time dependent
distributions of strings.  As an example we compute the rate of decay of
a small initial perturbation to an equilibrium string distribution.
In Section~\ref{seciii} we compare our results with calculations based
on the free string density of states and discuss further the physical
picture that emerges. Section~\ref{seciv} sets up similar Boltzmann
equations for long open strings. In this case, however, no long
string phase forms as the energy density is increased beyond
the Hagedorn energy density. The probability for string decay
increases as the length of string increases, while the probability
for rejoining decreases inversely with the volume, so that in the
thermodynamic limit long strings are suppressed in open string theory.
Section~\ref{secv} summarizes our conclusions.

\section{Boltzmann Equation for Long Closed Strings}
\label{secii}

In conventional statistical mechanics there are several ways to
obtain equilibrium distribution functions.  One is to start from the
single object density of states computed in the non-interacting theory
and using the equipartition theorem.  This is the route that has been taken
in much of the previous work on high-temperature string theory.  Another
method is to consider transport equations, which describe how energy and
other conserved quantities are redistributed in collisions among
constituents of the ensemble, and impose equilibrium conditions.

Consider, for example, the Boltzmann equation for a gas of interacting
massive particles
\begin{equation}
\label{boltz}
{ {\partial f(\vec v_1, t)} \over {\partial t}} =
\int d^D v_2 \int d\Omega \sigma(\Omega) |\vec v_1 - \vec v_2|
(f_2' f_1' - f_2 f_1) \,,
\end{equation}
where $ f(\vec v,t) d^D v$ is the number of particles per unit volume
lying in the velocity volume $d^D v$ about $\vec v$, and
$\sigma(\Omega)$ is the differential scattering cross section. This
equation describes the change in time of the distribution
function due to binary collisions of the form $\{ \vec v_1, \vec v_2\}
\to \{ {\vec v_1}', {\vec v_2}'\}$. The derivation of this
equation makes the crucial assumption that the velocity of a particle
is uncorrelated with its position, which is usually valid for
sufficiently dilute systems.  The equilibrium solution of the Boltzmann
equation (\ref{boltz}) is the Maxwell--Boltzmann distribution,
\begin{equation}
f(\vec v) \sim e^{-\beta m  v^2/2} \,,
\end{equation}
which describes a gas of particles in the canonical ensemble,
i.e. fixed volume and temperature.

We want to obtain an analogous equation for a gas of interacting
string loops.  In the limit of very small loops the extended nature
of strings becomes irrelevant and the result for massless
particles must be recovered.  At intermediate scales we expect the loop
equations of string theory to be complicated and we will
not attempt to derive them in full generality here.  Instead we
focus our attention on very long string loops, which satisfy rather
simple equations and whose distribution function can be
explicitly computed.  The limit of long strings is of considerable
interest since previous studies of the free string density of states
suggest that the microcanonical ensemble becomes dominated by long strings,
as the Hagedorn energy density is approached
\cite{fraut},\cite{carl},\cite{sund}-\cite{branden}.

A key observation is that when the string length becomes large compared
with the spatial size of the system the string will traverse the entire
volume many times over and, assuming a generic parametrization of the
string, two points on the string, which are separated
by a finite parameter distance, are found at uncorrelated
positions in the embedding space.  The Boltzmann equation for such
strings will only involve intrinsic properties of the
strings, such as their length, but not any details of their embedding.

A further simplification for long strings is that their energy is
dominated by the string tension so we can simply characterize the
single string states by their loop length.  It follows from the condition
(\ref{jeancond}) that the interaction energy is a small
fraction of the total energy of the system and at equilibrium that
fraction will not change with time. Thus we only include terms
in the Boltzmann equation that conserve total length, so that
to a good approximation the interactions will conserve energy.

The analog of the Boltzmann equation for long oriented closed strings
takes the following form:
\begin{eqnarray}
{{\partial n(\ell)} \over {\partial t}}
&=& {\kappa\over V} \Bigl\{ -{1\over 2}\ell^2 n(\ell)-
 \int_0^{\infty} d \ell' \ell' n(\ell') \ell n(\ell)
+ {1\over 2}\int_0^{\ell} d \ell'~
\ell' (\ell-\ell') n(\ell') n(\ell-\ell')
\nonumber\\
&&+
\int_{\ell}^{\infty} d \ell' \ell' n(\ell') \Bigr\},
\label{loope}
\end{eqnarray}
where $\kappa$ is some positive constant which depends
on the string coupling, and for convenience we have set $\alpha'=1$.
Here $n(\ell)$
is the average number of strings of length $\ell$.
The first term on the right hand side of (\ref{loope}) represents the
effect of a loop of length $\ell$ self-intersecting
and splitting into two smaller strings.
The factor of $\ell^2/V$ reflects the probability of finding two
bits of the same long string at the same point in the embedding space.
The second term comes from two strings of length $\ell$ and $\ell'$
joining into a single string of length $\ell+\ell'$.
The third term describes two smaller strings joining to form
a single string of length $\ell$.
The fourth term describes a long string of length $\ell'$ self-intersecting
and splitting to form a string of length $\ell$ and another string
of length $\ell'-\ell$.

All four terms on the right hand side of (\ref{loope}) thus involve
a three string interaction and we are ignoring contributions from
interactions of four or more strings, which are suppressed when the
string coupling is weak.  In general, the four terms carry different
phase space factors but in the long string limit these factors are the
same for all the terms, and are absorbed into the constant $\kappa$.
This is because the basic interaction is the same in all cases: two
short segments of string cut across each other and exchange ends at the
intersection point.  The interaction rate involves an average over
the relative orientation and momentum of the segments when they meet
but in the long string limit this average is not sensitive to the
overall string length nor to whether or not the segments belong to the
same long string before the interaction.

A similar Boltzmann equation holds for
unoriented closed strings, with extra factors of $1/2$ appearing
in front of the first and fourth terms on the right hand side of
(\ref{loope}). For simplicity, we will restrict our considerations
to oriented closed strings in the following, and note that
qualitatively similar conclusions hold for unoriented closed strings.

It should be noted that the Boltzmann equation (\ref{boltz}) is a
truncation of an exact set of equations known as the
Bogoliubov-Born-Green-Kirkwood-Yvon (BBGKY) hierarchy (see
\cite{lifpit} for details).  An analogous hierarchy governs the
exact statistical dynamics of strings but the string Boltzmann equation
(\ref{loope}) is sufficient for the purposes of this paper.

At equilibrium the average number of strings of a given length
remains constant,
\begin{equation}
{\partial n(\ell)\over \partial t} =0,
\end{equation}
and we therefore consider static solutions to the Boltzmann equation.
The Laplace transform of (\ref{loope}) then gives
\begin{equation}
{1\over 2}\partial_z \tilde w(z)
-{\bar L} \tilde w(z)
+ {1\over 2}\tilde w(z)^2
+{ {{\bar L}-\tilde w(z)}\over {z}} = 0,
\label{laploop}
\end{equation}
where we have defined $w(\ell)=\ell n(\ell)$ and $\bar L$ is the
average total length of string  in the ensemble, given by
$\bar L= \int_0^{\infty} d\ell w(\ell)= \tilde w(0)$.
The Laplace transformed equation is solved by
\begin{equation}
\tilde w(z) = {1 \over z+ 1/ {\bar L}} ,
\end{equation}
and taking the inverse Laplace transform gives
\begin{equation}
w(\ell) = e^{ - \ell / {\bar L}}  .
\end{equation}
The average number of strings as a function of length is then
\begin{equation}
n(\ell) = {e^{-\ell / {\bar L}}  \over \ell} .
\label{cstr}
\end{equation}
This is the analog of the Maxwell--Boltzmann distribution
for long fundamental strings, describing strings at fixed temperature
in the canonical ensemble.
Large energy fluctuations appear in the canonical ensemble as the
Hagedorn temperature is approached and at that point the canonical and
microcanonical ensembles are no longer equivalent.  It is
preferable to use the more fundamental microcanonical ensemble
to describe the physics.  In order to do that, we first obtain
the single string density of states from the canonical ensemble
distribution (\ref{cstr}) using
\begin{equation}
n(\ell) =  \omega(\sigma \ell) \sigma  e^{-\beta \sigma \ell},
\end{equation}
where $\sigma$ is the string tension and the energy of a single string
is to a good approximation given by $\varepsilon = \sigma \ell$.
The resulting density of states is
\begin{equation}
\omega(\varepsilon) = { {e^{\beta_H \varepsilon}} \over \varepsilon},
\label{bdos}
\end{equation}
where
\begin{equation}
\beta_H = \beta - {1 \over {\sigma {\bar L}}}
\label{cetemp}
\end{equation}
is to be identified with the inverse Hagedorn temperature. Note there is
no volume factor in (\ref{bdos}) so the density of long string states
is not an extensive quantity.  We will return to this point in the
following section.

Given the single string density of states (\ref{bdos}) one may obtain
the multi-string density of states,
\begin{equation}
\Omega(E) \approx e^{\beta_H E}.
\end{equation}
The single string distribution function for fixed total energy $E$
is then approximately
\begin{equation}
d(\varepsilon, E)  \approx {{\omega(\varepsilon) \Omega(E-\varepsilon)}
\over {\Omega(E) }} \approx {1\over \varepsilon}.
\label{dftn}
\end{equation}
This approximation is valid for $c < \varepsilon < E-c$ where $c$ is some
constant independent of $E$ \cite{deo}.

The number of long strings of a given length in the microcanonical
ensemble thus depends inversely on the length and the average total
number of long strings is $\log{E}$.  If we think of each string as a
collection of small bits, whose number is proportional to the string
length, then $\varepsilon d(\varepsilon,E)$
determines the relative probability that a given string bit
finds itself a part of a string of length $\ell=\varepsilon/\sigma$.
Since $\varepsilon d(\varepsilon,E)$ is independent of $\varepsilon$ the
string bit could belong to a string of any length with equal probability.
This is reasonable since a long string traverses the entire volume of
the system and its length can at any moment be modified by interactions
which occur far away from the given string bit.

One advantage of using a Boltzmann equation to describe
hot strings is that we can compute time dependent properties of the system.
As an example, consider introducing a small perturbation $\delta n(\ell)$
to the equilibrium distribution (\ref{cstr}) localized around some
length $\ell=\ell_0$.  Solving the linearized Boltzmann equation about
the solution (\ref{cstr}) one finds an exponential rate of decay of the
perturbation,
\begin{equation}
\delta n(\ell, t) = \delta n(\ell,0)
e^{-\kappa (\bar L+{1\over 2}\ell_0) \ell_0 t/V},
\end{equation}
confirming that the equilibrium distribution found above is stable.

\section{Comparison to Free String Calculations}
\label{seciii}

The above results can be compared to those obtained by direct
counting of free string states but that requires some care.
Weak string interactions must be introduced to allow the system
to explore phase space, so that the equipartition theorem holds.
As emphasized above, the Jeans instability then forces
one to consider systems with finite size.  In order for the
calculation based on the free density of states to be fully
consistent the target space should be taken to be compact.
A thermodynamic limit may be taken at the end of the day provided
the string coupling is scaled to zero sufficiently rapidly.

It has been shown by Brandenberger and Vafa \cite{branden} that,
remarkably, the free string density of states in an arbitrary compact
target space takes a universal form in the high energy limit.
This may be seen by considering the one-loop partition
function for strings on a compact target space at finite
temperature $T=1/\beta$. The time is now regarded as a
compact Euclidean direction with period $\beta$. As the
Hagedorn temperature is approached the partition function
begins to diverge as the state which winds once around the
time direction becomes massless \cite{sath}-\cite{obrien}.
One finds that the expectation value of the energy diverges as
\begin{equation}
E \approx - {\partial \over {\partial \beta}} \int^{\infty}
{{d\tau}\over \tau} e^{-(\beta-\beta_H) \tau}
= {1 \over {\beta-\beta_H}} + \cdots ,
\end{equation}
from which one may deduce that the asymptotic high energy
density of states is given by (\ref{intdos}) \cite{branden}.
Thus free string calculations agree with results obtained from the
Boltzmann equation, provided the thermodynamic limit is taken in the
correct way, i.e. taking the limit of infinite volume only at the
very end.

The single string density of states for closed strings in a large
but finite volume $V$ has the form (\ref{freedos}) as long as the
string length remains small compared to the size of the system.
It then crosses over to (\ref{intdos}) in the high-energy limit
once the string traverses the entire volume.  The two expressions
for the density of states differ by factor of
$V/\varepsilon^{D\over 2}$, which can be understood by the following
heuristic argument due to Polchinski \cite{pol}.  The shape of a highly
energetic string is approximated by a random walk.
If the string is not too long the random walk occupies a volume of
order $\ell^{D\over 2}\sim (\varepsilon/\sigma)^{D\over 2}$, and the
density of states contains a factor of $V/\ell^{D\over 2}$ from
the translation zero mode.  If, on the other hand, the string is so
long that $\ell > V$ then the random walk fills the entire system
volume so that this zero mode factor is absent and the density of
states takes its high energy form.

The two expressions (\ref{freedos}) and (\ref{intdos}) for the free
string density of states lead to very different thermodynamic
behavior.  If one uses the density of states in a non-compact space
(\ref{freedos}) one is led to conclude that there is a long string
phase dominated by a single string (or very few)
\cite{fraut},\cite{carl},\cite{sund}-\cite{deo}, and further that the
specific heat derived from (\ref{freedos}) is negative, indicating an
instability.
The calculation which leads to the single string dominance is,
however, inherently inconsistent, since, as stated above, once one
allows arbitrarily weak interactions then the Jeans instability sets in
and one has to consider a finite volume target space.

In order to determine the sign of the specific heat
in the long string phase dictated by (\ref{intdos}) it is
necessary to calculate higher order corrections to the
density of states.  Brandenberger and Vafa \cite{branden} showed that
these corrections lead to an entropy $S(E)$ in the microcanonical
ensemble of the form
\begin{equation}
S= \beta_H E - {C \over E} + {\rm const.}
\end{equation}
where $C$ is a positive constant. This implies
\begin{equation}
\beta = \biggl( {{\partial S} \over {\partial E}} \biggr)_V = \beta_H +
{C \over {E^2}},
\label{metemp}
\end{equation}
and one finds that the specific heat is positive. At this point
one might be surprised at the apparent discrepancy between equations
(\ref{metemp}) and (\ref{cetemp}). It is a consequence of the
inequivalence of the canonical and microcanonical ensembles as
the Hagedorn temperature is approached as discussed above.

To sum up, the single long string phase is unstable
when interactions are included. Instead, the stable long
string phase involves a distribution of long strings which
may split and join as described in the previous section.

\section{Boltzmann Equation for Long Open Strings}
\label{seciv}

One may apply similar arguments to set up a Boltzmann
equation describing long open strings.  In this case we assume a
string may split with a constant probability per unit length
\cite{polchb}.
Unlike in the closed string case, this does not require
self-intersection. When open strings join,  the ends of two
strings must collide, the likelihood of which falls inversely
with the volume of the system \cite{salom}.
The Boltzmann equation for long oriented open strings is then
\begin{eqnarray}
{1\over {\kappa}}{ {\partial n(\ell)} \over {\partial t}}
&=& -\ell n(\ell)-
  {a\over V}\int_0^{\infty} d \ell' n(\ell')  n(\ell) +
{a\over 2V}\int_0^{\ell} d \ell'~
n(\ell') n(\ell-\ell')  \nonumber\\
&&+
2\int_{\ell}^{\infty} d \ell'  n(\ell').
\label{loopeo}
\end{eqnarray}
Here $\kappa$ is a positive constant that depends on the string
coupling and $a$ is a positive constant related to the string
joining probability.

To find the static solutions, it is again convenient to take a Laplace
transform to get
\begin{equation}
 \tilde n'(z)
-{a\over V}\bar N \tilde n(z)
+ {a\over 2V}\tilde n(z)^2 +
{2\over z} \bigl(\bar N-\tilde n(z)\bigr) = 0.
\label{laploopo}
\end{equation}
Here $\bar N$ is the average total number of open strings.  The solution is
\begin{equation}
\tilde n(z) = {2V \over a(z+ 2V / \bar N a) },
\end{equation}
and taking the inverse Laplace transform gives
\begin{equation}
n(\ell) = {2V\over a} e^{ - 2V \ell / \bar N a }.
\label{osdos}
\end{equation}
This then is the analog of the Maxwell--Boltzmann distribution
for long open strings in the canonical ensemble.
Reading off the density of states, one obtains
\begin{equation}
\omega( \varepsilon) \approx {2V\over a\sigma} e^{\beta_H \varepsilon}.
\end{equation}
Unlike for closed strings, the density of states of open strings is an
extensive quantity.

Now consider the open string distribution function in the microcanonical
ensemble.  The multi-string density of states, in a saddle point
approximation valid in the limit, $V\to \infty$ with $E/V$ fixed, is
\begin{equation}
\Omega(E) \approx
\exp{\left\{\sqrt{8EV\over a\sigma}+\beta_H E\right\}},
\end{equation}
and for open strings carrying a finite fraction of the total energy of
the system the string distribution function,
\begin{eqnarray}
d(\varepsilon, E) &=& {{\omega(\varepsilon) \Omega(E-\varepsilon)}
\over {\Omega(E)}}, \nonumber\\
&\sim &
\exp{\left\{-\sqrt{8EV\over a\sigma}
(1-\sqrt{1-\varepsilon/E})\right\}},
\end{eqnarray}
is strongly suppressed in this limit.  Thus we conclude that no long open
string phase exists in the thermodynamic limit in agreement with the
discrete model results of Salomonson and Skagerstam \cite{salom}.
Similarly, if closed strings are allowed to decay into open
strings, no long closed string phase will form either.

\section{Discussion}
\label{secv}

In this paper we presented a Boltzmann equation describing the
evolution of long fundamental strings.  The equilibrium solution
is easily found and describes a
distribution where the available energy, $E$, is shared evenly
between strings of different length and the total number of long
strings is of order $\log{E}$.  Our results agree with those of
Salomonson and Skagerstam \cite{salom} who studied a discrete
model of weakly interacting strings.  This is not surprising since
for the most part they made the same physical assumptions as we
have.  The advantage of our continuum approach is that it neatly
summarizes the combinatorics of the discrete model and yields
the equilibrium distribution with minimal effort.  The Boltzmann
equation can also be utilized to discuss non-equilibrium
behavior.

The equilibrium solution of the Boltzmann equation corresponds to
a single string density of states which agrees with that obtained
by direct computation, provided care is taken to work on a
compact target space \cite{salom},\cite{branden}.  It is
reassuring that these two different approaches to the statistical
mechanics of strings are mutually consistent.

The physical picture that emerges from these considerations is
as follows.  Imagine a gas of weakly interacting strings in a
space of large but finite volume as the energy density is
slowly increased.  At first the strings will  predominantly be
small and their behavior will be governed by some low energy
effective field theory.  As the Hagedorn energy density is
approached long strings begin to form.  Eventually the ensemble
will be dominated by the long string configurations described in
this paper although there presumably remains a small component
of short strings which behaves as a thermal gas of particles in
equilibrium with the long strings.

According to equation (\ref{metemp}) the Hagedorn temperature
is a limiting temperature in the microcanonical ensemble.  As
further energy is pumped into the system most of it is spent on
forming long strings rather than increasing the temperature.
This behavior is reminiscent of a first order phase transition
with a large latent heat.  Our approximations break
down when the energy density becomes too large.  At
$\rho \sim 1/g^2\alpha'$ the string theory becomes strongly
coupled but even before that the Jeans instability undermines
the thermal ensemble.  Once the inequality in (\ref{jeancond})
is violated the system becomes unstable to gravitational
collapse.

The Boltzmann equation provides a starting point for the study of
non-equilibrium thermodynamics of fundamental strings.
An important problem for future work is to generalize to
non-trivial metric and dilaton backgrounds in order to get a
handle on the gravitational collapse of string distributions.
Such a generalization would also be relevant to the description of
cosmic strings in the early universe.

\vskip 10pt
{\bf Acknowledgements:}

\noindent
We are grateful to S.~Giddings, S.~Ramaswamy, A. Strominger
and especially J.~Polchinski for useful discussions. We also thank
P.~Krapivsky for helpful comments on a previous version of the
manuscript.  This work was supported in part by NSF~Grants
PHY91-16964 and PHY-89-04035.


\begin{thebibliography}{99}

\bibitem{hage} R. Hagedorn, Nuovo Cim. Suppl. {\bf 3} (1965) 147.

\bibitem{huang} K. Huang and S. Weinberg, Phys. Rev. Lett. {\bf 25}
(1970)
895;
S. Fubini and G. Veneziano, Nuovo Cim. {\bf 64A} (1969) 1640.

\bibitem{fraut} S. Frautschi, Phys. Rev. {\bf D3} (1971) 2821.

\bibitem{carl} R.D. Carlitz, Phys. Rev. {\bf D5} (1972) 3231.

\bibitem{atick} J.J. Atick and E. Witten,
Nucl. Phys. {\bf B310} (1988) 291.

\bibitem{gsw} See {\it e.g.} M.~Green, J.~Schwarz, and E.~Witten,
{\it Superstring Theory, Vol.~1}, Cambridge Univ.~Press, 1987.

\bibitem{sund} B. Sundborg, Nucl. Phys. {\bf B254} (1985) 538.

\bibitem{mitch} D. Mitchell and N. Turok, Phys. Rev. Lett. {\bf 58}
(1987) 1577; Nucl. Phys. {\bf B294} (1987) 1138.

\bibitem{bowgid}
M.J. Bowick and L.C.R. Wijewardhana, Phys. Rev. Lett.
{\bf 54} (1985) 2485;
M.J. Bowick and S.B. Giddings, Nucl. Phys. B325 (1989) 631;
S.B. Giddings, Phys. Lett. {\bf 226B} (1989) 55.

\bibitem{deo} N. Deo, S. Jain and C-I Tan, Phys. Rev. {\bf D40} (1989) 2626;
Phys. Lett. {\bf 220B} (1989) 125;  N.~Deo, S.~Jain, O.~Narayan and
C-I~Tan, Phys. Rev. {\bf D45} (1992) 3641.

\bibitem{salom} P. Salomonson and B. Skagerstam, Nucl. Phys. {\bf B268}
 (1986) 349; Physica {\bf A158} (1989) 499.

\bibitem{branden} R. Brandenberger and C. Vafa, Nucl. Phys. {\bf B316}
(1989) 391.

\bibitem{lifpit} E. Lifshitz and L. Pitaevskii, {\it Physical Kinetics},
Pergamon Press, 1981.

\bibitem{sath} B. Sathiapalan, Phys. Rev. {\bf D35} (1987) 3277.

\bibitem{kogan} Ya.I. Kogan, JETP Lett. {\bf 45} (1987) 709.

\bibitem{obrien} K.H. O'Brien and C-I Tan, Phys. Rev. {\bf D36} (1987)
1184.

\bibitem{pol} J. Polchinski, private communication.

\bibitem{polchb} J. Dai and J. Polchinski, Phys. Lett. {\bf 220B} (1989) 387.

\end{thebibliography}
\end{document}